# Hysteretic magnetic pinning and reversible resistance switching in High-Tc superconductor/ferromagnet multilayers

C. Visani, P. J. Metaxas, A. Collaudin, B. Calvet, R. Bernard, J. Briatico, C. Deranlot, K. Bouzehouane and J. E. Villegas[*]

*Unité Mixte de Physique CNRS/Thales, 1 ave. A. Fresnel, 91767 Palaiseau, and Université Paris Sud 11, 91405 Orsay, France*

We study a high-$T_C$ superconducting ($YBa_2Cu_3O_{7-\delta}$) / ferromagnetic (Co/Pt multilayer) hybrid which exhibits resistance switching driven by the magnetic history: depending on the direction of the external field, a pronounced decrease or increase of the mixed-state resistance is observed as magnetization reversal occurs within the Co/Pt multilayer. We demonstrate that stray magnetic fields cause these effects via i) creation of vortices/antivortices and ii) magnetostatic pinning of vortices that are induced by the external field.

*corresponding author: **javier.villegas@thalesgroup.com**

Hybrid structures combining dissimilar materials often exhibit unusual properties, which result from the interaction between competing order parameters. Understanding the underlying mechanisms allows one to tailor materials with novel functionalities. In this sense, Superconductor/Ferromagnet (S/F) hybrids are paradigmatic. Their transport properties result from the interplay of a variety of nanoscale physical mechanisms: conventional [1] or unconventional proximity effects [2], injection of spin-polarized currents [3], and effects related to stray magnetic fields which emanate from the F [4,5]. More work has been devoted to S/F hybrids with conventional low-$T_C$ superconductors (see e.g. [6,7,8]) than with high-$T_C$ superconductors (see a review in [9]).

Much attention has been paid to S/F systems in which changes in the F magnetic state induce a resistance switching in the S. For example, in certain F/S/F trilayers, the resistance near the critical temperature $T_C$ depends on whether the F layers' magnetizations are parallel or antiparallel [10-20]. This effect has been explained in terms of various mechanisms that produce a shift of the critical temperature $\Delta T_C$ depending upon the configuration of the F layers: Cooper-pair breaking either i) due to the exchange field induced in the S [10,12,18,19,23] or ii) due to the accumulation of spin polarized quasiparticles [11,13,16,22] and iii) Cooper-pair formation due to crossed Andreev reflection [24]. On the other hand, various low-$T_C$ S/F systems show resistance switching effects produced by stray magnetic fields generated by the F layers' domain structure [5,14,15,17,20]. This possibility has also been contemplated for high-$T_C$ systems [21].

In this article, we report on reversible resistance switching effects caused by stray magnetic fields in high-$T_C$ S/F hybrids. Specifically, we show that the tunable domain structure of a F (Co/Pt superlattice) causes hysteretic magneto-transport in a high-$T_C$ superconducting YBa$_2$Cu$_3$O$_{7-\delta}$ (YBCO) film via (i) the creation of vortices/antivortices and (ii) the pinning of the vortices that are induced by the external applied field. While (i) induces an increase of the mixed-state resistance, (ii) induces a decrease. These changes depend upon both the magnetic history and the direction of the external applied magnetic field. Previously, similar experiments in which a low-$T_C$ S (Nb) was used were interpreted in terms of $T_C$ variations caused by Cooper-pair breaking, and stray magnetic field effects were dismissed [19]. However, our combination of Magnetic Force Microscopy (MFM), Anomalous Hall Effect (AHE) [25] measurements, current dependent magneto-transport measurements and magnetostatic calculations allow us to unambiguously connect the resistance switching observed here to vortex dynamics effects induced by the stray fields. Notably, we can correlate the magnetic history dependent pinning of vortices [26] with the varying F domain structure.

For the fabrication of the S/F structure, a 20 nm thick c-axis YBCO films was grown on (001) SrTiO$_3$ using pulsed laser deposition (KrF excimer). Atomic force microscopy (not shown) revealed a flat film surface (roughness ~ 1 nm). A Pt$_{10\text{ nm}}$/(Co$_{0.6\text{ nm}}$/Pt$_{1\text{ nm}}$)$_5$/Pt$_{5\text{ nm}}$ multilayer (with Pt$_{10\text{ nm}}$ the buffer layer) was subsequently sputtered *ex situ* at room temperature directly on top the YBCO. A bridge (250 μm long × 40 μm wide) for standard four-probe resistance and Hall (voltage perpendicular to current) measurements was optically lithographed and ion etched. Magneto-transport measurements were carried out in a He flow cryostat equipped with a rotatable sample holder and a 5.5 kOe electromagnet. The zero-field critical temperature obtained from R(T) measurements (the onset of the superconducting transition is defined as the temperature at which the resistance falls to 90% of the normal-state value) was $T_C = 72$ K for the S/F sample and $T_C = 82$ K for a single 20 nm thick YBCO film (with no F on top). The latter was used as reference sample. The "zero-resistance" state (as defined by a measured V~10$^{-7}$ V for an injected current I=1 μA) was achieved at T~66 K for both samples. Note that shunting of the injected currents across the Co/Pt multilayer occurs until the YBCO layer resistance has significantly decreased with respect to its normal-state value, which is evidenced by the very different normal-state sheet resistances of the S/F sample $R_s(90K) \sim 6~\Omega$ and the single YBCO film $R_s(90K) \sim 320~\Omega$.

Figure 1 (a) shows the central results of this paper. This figure displays the mixed-state resistance *vs.* field of the S/F sample for various angles $\theta$ between $H$ and the c-axis (see inset). Measurements were carried out in constant Lorentz force geometry (current $J \perp H$). For each $\theta$, $R(H)$ was measured as $H$ was swept from positive to negative ($R_{DEC}(H)$, black line) and *vice versa* ($R_{INC}(H)$, red dashed line). After demagnetization [27] (with $H$ applied at $\theta$ from the c-axis), a third measurement was performed in which $H$ was swept to positive fields from $H=0$ (the "virgin" $R_{VIR}(H)$, blue circles). *The magneto-transport is hysteretic for all $\theta$*: depending upon the magnetic history, a switching between high and low resistance states is observed within a range of positive/negative fields, so that $R_{DEC}(H)$ and $R_{INC}(H)$ form two "lobes" which appear symmetrically around $H=0$. *However, distinct behavior is observed for $\theta=90°$ and $\theta<90°$. Moreover, the zero-field resistance strongly depends on the magnetic history and $\theta$*.

For $\theta<90°$, $R_{DEC}(H)$ and $R_{INC}(H)$ coincide except for a range of negative (positive) $H$ within which $R_{DEC}(H) < R_{INC}(H)$ ($R_{DEC}(H) > R_{INC}(H)$). As $\theta$ is increased, the background magneto-resistance diminishes, and the "lobes" widen. Regarding the virgin $R_{VR}(H)$ (blue circles, the curve for $\theta=70°$ is representative of all $\theta<90°$ curves), we observe that $R_{VIR}(0)$ is about one order of magnitude higher than $R_{DEC}(0) = R_{INC}(0)$. However, $R_{VIR}(H)$ eventually crosses under and matches $R_{INC}(H)$ as $H$ is increased.

The behavior for $\theta=90°$ is very different. The background magneto-resistance is nearly constant, and at high field it is over two orders of magnitude lower than for $\theta<90°$. The increasing and decreasing field branches do not coincide for any $|H| \leq \sim 3$ kOe except for $H=0$, where they cross. *Note that the resistance switching is reversed as compared to the* curves for $\theta < 90^o$: here $R_{DEC}(H) > R_{INC}(H)$ for negative $H$, and vice versa for positive $H$. The virgin $R_{VIR}(H)$ (blue dots) is higher than $R_{INC}(H)$ and $R_{DEC}(H)$ for all $|H|<\sim 3$ kOe, and matches the high-field resistance background above that field.

Figure 1 (b) shows the magneto-transport of a single 20 nm thick YBCO film, which, as expected, does not exhibit hysteresis. Comparison of the curves in Figs. 1 (a) and (b) allow us to separate the effects of $H$ (background magneto-resistance induced by flux dynamics in the YBCO film) from those caused by the presence of the Co/Pt multilayer.

In Figure 2, we demonstrate the correlation between the multilayer's magnetic state and the hysteretic magneto-resistance of the S/F sample. For each $\theta$, the top panel depicts *the percent resistance switching with respect to the background magneto-resistance*, $\Delta R(H)$, and the lower panel shows the normalized *net* perpendicular component of the multilayer's magnetization $M_\perp(H)/M_S$ (with $M_S$ the saturation magnetization).

$M_\perp(H)/M_S$ was obtained from AHE [25] measurements at T=100 K (i.e. in the normal state). Note that, due to the much lower normal-state sheet resistance of the Co/Pt multilayer as compared to the YBCO layer, nearly 98% the injected current flows within the former. The measured Hall voltage contains both the ordinary component (OHE) and the anomalous one (AHE) characteristic of ferromagnetic systems [25]. Unlike AHE, OHE *is not hysteretic and is directly proportional to the component of $H$ perpendicular to the film plane*. Discrimination between OHE and AHE is therefore straightforward from a set of Hall effect measurements for different $\theta$. This allows for the extraction of the AHE resistance, which is directly proportional [25] to the *net* out-of-plane component of the magnetization $M_\perp(H)/M_S$ (normalized).

For $\theta=0$ [Fig. 2 (a)], $M_\perp(H)$ is as expected for a system with strong perpendicular magnetic anisotropy, in which magnetic reversal occurs via the nucleation and subsequent growth of domains (eg. [28]). *Around the coercive field ($M_\perp \sim 0$) up/down magnetized domains are formed which are expected to be comparable to those in the MFM images of Fig. 4.* For $0<\theta<90°$ [Fig. 2 (b) and (c)], the remanent perpendicular magnetization $|M_\perp(0)| \sim M_S$ !suggests that the magnetization coherently rotates out-of-plane as $|H|$ is decreased to zero following saturation in high field. As $H$ is reversed, $M_\perp$ eventually switches via nucleation and growth of reverse domains under the action of the reverse out-of-plane component of $H$. Finally, the magnetization again coherently rotates towards the $H$ direction as the field magnitude is further increased.

For $\theta=90°$, $M_\perp(H)$ [Fig. 2 (d)] and MFM [Fig 4 (a)] imply the following magnetic reversal mechanism. At high in-plane $|H|$, the magnetization lies essentially in the film plane. As $|H|$ is reduced from its high value to zero, the magnetization gradually rotates out of the film plane under the influence of the perpendicular anisotropy and breaks up into domains. The remanent domain structure is shown in Fig. 4 (a). In principle, one expects an equal number of "up" and "down" domains, resulting in $M_\perp=0$. Experimentally however, we do see some hysteretic effects in $M_\perp(H)/M_S$ [Fig. 2(d)] which is suggestive of a slight misalignment of the field and the Co/Pt multilayer. This will slightly favor the positively (negatively) magnetized domains following positive (negative) in-plane saturation. This biasing effect is rather weak though, yielding a maximum value of $|M_\perp| \sim 0.08 M_S$. Unexpectedly, this maximum value is obtained at $|H| \sim 0.5$ kOe and not at zero field, at which point no canting is to be expected, thereby being the field at which the maximum $|M_\perp|$ should occur. The explanation for this discrepancy is not clear at this time but may be related to the existence of a small in-plane magnetization component at remanence. As $H$ is further reversed beyond $|H| \sim 0.5$ kOe, the magnetization gradually rotates back in-plane and the domains are annihilated.

We detail now the correlation between the resistance switching $\Delta R(H)$ and the magnetization reversal for different $\theta$. For $\theta<90º$, we defined $\Delta R(H) \equiv (R_{DEC}(H) - R_{INC}(H))/R_{INC}(H)$ for $H<0$ ($R_{INC}$ and $R_{DEC}$ are permuted in the formula for $H>0$). For $\theta=90º$, given the nearly constant background resistance (see figure 1 (b)), we defined $\Delta R(H) \equiv (R_{DEC}(H) - R_{DEC}(4\,kOe))/R_{DEC}(4\,kOe)$ for the decreasing field branch (black), and used the same formula with $R_{INC}$ for the increasing field branch (red). For $\theta<90º$ [top panels, Fig. 2 (a)-(c)] $\Delta R(H)$ dips are observed within the field range in which $M_\perp$ reverses, with $|\Delta R(H)|$ at its maximum for $M_\perp \sim 0$. As we change $\theta$ from 0 to 80º the amplitude of $\Delta R(H)$ gradually increases, and the dips become smoother and wider, since the reversal of $M_\perp$ becomes less abrupt. For $\theta=90º$, [Fig. 2 (d)], a direct correlation between $\Delta R(H)$ and $|M_\perp(H)|$ is observed, and the maximum $\Delta R(H)$ (~90%) occurs at the field $|H| \sim 0.5$ kOe at which the $M_\perp(H)/M_S$ is maximum. Note also that the sign of $\Delta R(H)$ is reversed as compared to the curves for $\theta<90º$. The most important conclusion of Figure 2 is that, depending on whether $H$ is applied out-of-plane ($\theta<90º$) or in-plane ($\theta=90º$) the presence of a structure of up/down magnetized domains respectively produces a decrease or a (relatively larger) increase of the mixed-state resistance. This qualitative behavior is observed for all T between $0.80\,T_C$ and $0.99\,T_C$.

The current and temperature dependences of $\Delta R$ are shown in Figure 3. In Fig. 3 (a) we have plotted $E(J)$ at different temperatures (see labels) in $H=1$ kOe (~ the field at which the maximum $\Delta R$ is observed, see Fig. 2 (c)). The black solid curves correspond to the case in which the multilayer's magnetization is homogeneous (which is after $H$ is swept from +3.5 kOe to +1 kOe). The red dashed curves correspond to the case in which the multilayer presents a structure of up/down magnetized domains (after $H$ is swept from –3.5 kOe to +1 kOe). It can be seen that the red dashed curves are shifted to the right with respect to the black ones, which implies a lower resistance when the magnetic domains structure is present. *Note that this effect spans over a wide temperature range, both above and below the irreversibility line* (indicated by the black dashed straight line). Figure 3 (b) shows current dependence of the resistance switching $\Delta R$, which was calculated from the set of $E(J)$ shown in Fig. 3(a). For a given temperature, $|\Delta R|$ decreases as $J$ increases. For a given current, $|\Delta R|$ increases as the temperature decreases. The inset of Figure 3 (b) displays the critical current enhancement $\Delta J_C$ induced by the domain structure, for several temperatures below the irreversibility line. To obtain $\Delta J_C$ we measured *two* $E(J)$ for each field value, $H$: one following the application of +3.5 kOe, and the other following the application of –3.5 kOe. $J_C$ from each $E(J)$ was calculated with the criterion $E_C = 2 \cdot 10^{-4}$ V cm$^{-1}$. The maximum $\Delta J_C \sim 2$ kA cm$^{-2}$ is observed for $H \sim 1$ kOe. This corresponds to the critical current enhancement due to the presence of the domain structure as compared to the case in which the magnetization is homogeneous.

Figures 3 (c) and (d) show the same measurements as in (a) and (b), but with the field applied in-plane ($\theta$=90º). The black solid curves are measured for $H$ =-4 kOe, for which the magnetic superlattice is homogeneously magnetized essentially *in-plane*. The red dashed curves are measured for $H$ =0.5 kOe, for which the multilayer presents a structure of up/down magnetized domains similar to those in Figure 4 (a). Note that here the red dashed curves are shifted to the left with respect to the black ones, which implies a higher resistance when the magnetic domain structure is present. Note that for $\theta$=90º the resistance switching effects also span over a wide temperature range. As shown in Figure 3 (d), $\Delta R$ can reach up to ~ 200% at low temperatures.

The current and temperature dependences described above suggest that the resistance switching effects are connected to flux dynamics phenomena, as they are observed in a wide range of temperatures, both in the linear and non-linear regimes of $E(J)$. In what follows, we discuss in detail the origin of the observed behavior.

Hysteretic resistance switching effects have been previously observed in low-$T_C$ S/F multilayers with perpendicular magnetic anisotropy [19]. Those effects were explained by a shift of the critical temperature $\Delta T_C$ that was produced by the ferromagnet's *exchange* field. In this scenario, the strongest depression of superconductivity (i.e. that which results in the highest resistance state) corresponds to the case in which the ferromagnet is homogeneously magnetized. Lower resistance states are observed as the magnetization breaks into domains and superconductivity nucleates underneath the domain walls where the Cooper pairs experience weaker exchange field [19]. While the present results for $\theta < 90$ might *a priori* be understood within this picture, the behavior for $\theta = 90$ (Figs. 1 (a) and 2 (d)) rules out this possibility. Contrary to what is expected in the exchange field scenario [19], here the lowest resistance state is observed when the Co/Pt multilayer is homogenously magnetized (under the application of a ~4 kOe in-plane field), *and the presence of magnetic domains leads to a resistance increase* (see curve for $\theta = 90$). We argue below that the hysteretic magneto-transport observed in the present experiments is caused by the stray magnetic fields from the Co/Pt multilayer's magnetic domains. We show that this mechanism allows us to understand the observed behavior in the entire experimentally probed range of field angles. We consider only the *perpendicular* component of the stray field: the large anisotropy and the nearly constant background magneto-resistance for $H$ in-plane (see $\theta$ =90º in Fig. 1 (b)) imply that the effects of parallel magnetic fields are negligible as compared to those of perpendicular ones, as expected for thin YBCO films [29].

We first explain the very different zero-field *remanent* resistances measured after different magnetic preparations. *We find that the presence of flux quanta (vortices) induced by the corresponding magnetic domain structures accounts for the observed behavior.* In Figs. 4 (a), (b) and (c) we respectively show room-temperature MFM images of the remanent domain structures following I) the application and withdrawal of a saturating field at $\theta=90º$, II) demagnetization of the F [27] at $\theta=90º$ and III) demagnetization of the F at $\theta=0º$. A maze-like structure of up/down magnetized domains is observed in all cases, but the domain widths are significantly different for each of them. The corresponding stray field profiles $H_S(x)$ (displayed in Figs. 4 (d), (e) and (f) respectively) were numerically calculated from the digitized MFM images [30]. As the average domain width $\langle w \rangle$ increases (~250, ~330 nm and ~660 nm for states I, III and III), the spatially averaged magnitude of the stray field $\langle |H_S| \rangle$ decreases (~230, ~210, and ~150 Oe respectively). Note that $\langle |H_S| \rangle \sim 0$ when the multilayer is fully magnetized perpendicular to the film plane ($M_\perp \sim M_S$, as occurs after applying and then removing a saturating $H$ with $0 \leq \theta < 80º$). For states I, II, and III $H_S(x)$ oscillates from positive to negative, mimicking the structure of magnetic domains. As expected from theory [31,32] this will induce vortices and antivortices (vortices of opposite polarity) in the YBCO film. However, due to flux quantization, the creation of vortices by the stray field will depend on the stray field magnitude and the characteristic magnetic domain size. In order to quantify this, and to accordingly characterize the domain structures in the different magnetic states, we use the parameter $\phi/\phi_0 \equiv \pi \langle |H_S| \rangle \langle w \rangle^2 / 4\phi_0$ (with $\phi_0 = 2.07 \ 10^{-15}$ Wb the flux quantum). As defined, $\phi/\phi_0$ is the number of flux quanta induced by the *average* stray field within a circular area of diameter $\langle w \rangle$ underneath a magnetic domain [e.g. the (blue) circle in Fig. 4 (c)]. The definition of $\phi/\phi_0$ is motivated by the fact that domains have irregular, somewhat elongated shapes [Fig. 4 (a)-(c)]: considering that vortices are naturally isotropic in the film plane, one expects them to be formed under regular areas through which the net magnetic flux exceeds the flux quantum, and not underneath too narrow domains (regardless of their length). Experimentally, we find [Fig. 4 (g)] that the (temperature-dependent) zero-field resistances measured for the different magnetic states scale with $\phi/\phi_0$. This can be interpreted as follows. $\phi = 0$ corresponds to the state in which $M_\perp/M_S \sim 1$ and the stray field is essentially zero. In this case, essentially no vortices/antivortices are induced by the ferromagnet in the YBCO film and, consequently, the lowest zero-field resistance is observed. For the states I-III, the resistance increases as average domain width and $\phi/\phi_0$ do, because an increasing number of vortices/antivortices are induced by the stray field. Although these are localized underneath up/down magnetized domains, electrical resistance arises due to channeling of the vortices along the stripe-like domains parallel to the Lorentz force [33], thermal fluctuations, and vortex-loop excitations induced by the current [34]. $\phi/\phi_0 \ll 1$ indicates that many of the magnetic domains are too narrow to create a single vortex, so that vortices/antivortices are induced only by a fraction of them (larger than average). As $\phi/\phi_0$ increases, more vortices are induced, which yields a higher resistance. For the state III, in which the magnetic domains are the largest, $\phi/\phi_0 \gg 1$ and we expect stray-field induced vortices/antivortices underneath a majority of the domains. Note that the resistance dependence on $\phi/\phi_0$ tends to saturate above $\phi/\phi_0 = 1$. This is consistent with the above description: once all of the magnetic domains are large enough to create at least one single vortex ($\phi/\phi_0 > 1$), further increase of their size will not result in a larger density of vortices/antivortices across the film, and therefore no significant resistance increase is expected.

From the above, we can understand $R(H)$ for $\theta = 90º$. The in-plane $H$ has little effect on the YBCO thin film, but it changes the magnetic structure of the Co/Pt multilayer as it is cycled, indirectly producing the behaviors of Fig. 1 and 2. When $H$ is decreased to zero, $R_{DEC}(H)$ gradually increases as the multilayer breaks into domains, the magnetization rotates up/down out-of-plane, and vortices/antivortices are induced in the YBCO film. The maximum resistance is observed around $|H| \sim 0.5$ kOe, when the out of plane field component is at its maximum, and therefore $\langle|H_S|\rangle$ and the number of vortices/antivortices are maximum. Further decrease of $H$ leads to a gradual decrease of the resistance, as the magnetization again becomes uniform (thereby diminishing the stray field) and rotates back in-plane. The same description applies to $R_{INC}(H)$ as field is swept from negative to positive. A higher resistance is exhibited by $R_{VIR}(H)$ (blue circles in Fig 2), because the larger domains obtained after demagnetization produce more flux quanta [compare Figs. 4(a) and 4(b)].

To explain the behavior for $\theta < 90º$, we must also consider the vortices induced by the external field $H$. When the latter is sufficiently intense, vortices induced by $H$ outnumber the vortices/antivortices induced by the stray field, and become dominant. In this situation the structure of up/down magnetized domains produces magnetic pinning of vortices, as recently found in Nb/Co-Pt multilayers [34]. *We argue that this mechanism allows to explain why dips in $\Delta R(H)$ are observed [Figs. 2 (a)-(c)].* At low $H$ and in the presence of the maze-like domain structure (see e.g. $R_{VIR}(H)$, Fig 1 (a)), vortices/antivortices induce dissipation in the way described above. As $H$ is increased, it suppresses the antivortices under the magnetic domains having opposite polarity to it, and adds extra vortices. As theoretically shown [36] and experimentally observed in low-$T_C$ systems [37], the latter will be attracted to (repelled from) magnetic domains having the same (opposite) polarity, because this reduces the system's magnetostatic energy by a factor $U_m \sim \phi_0 \langle|H_S|\rangle$. This results in an enhancement of vortex pinning (and consequently in a resistance decrease) as compared to the case in which the multilayer's magnetization is uniform (and $H_S \sim 0$), which produces the dips of $\Delta R(H)$ [Figs. 2 (a)-(c)]. The maximum critical current enhancement $\Delta J_C \sim 2$ kA cm$^{-2}$ (see inset in Figure 3 (b)) is notably smaller than $\Delta J_C \sim 10^3$ kA cm$^{-2}$ estimated by equating the magnetic pinning force $U_m/\langle w \rangle$ to the Lorentz force $J\phi_0$ [36]. One may argue that this disagreement arises from a modified $\langle w \rangle$ in the superconducting state. However, in order to account for this discrepancy, the domains would need to be around two orders of magnitude larger in the superconducting state than at room temperature. More probably, that disagreement arises from the incommensurability between the vortex-lattice and the domain structure in the field range where dissipation is dominated by $H$. The expected distance between vortices $d \sim (\phi_0/H\cos(\theta))^{1/2} \sim 250$ nm is smaller than average distance $\sim 2\langle w \rangle$ between "pinning" domains (those having the same polarity as vortices). Therefore, matching of the vortex-lattice to the domains structure implies a large cost of elastic energy, and strongly reduces the available net magnetic pinning energy per vortex. In addition, the average domain size $\langle w \rangle \sim 0.5$-$0.6$ μm is comparable but shorter than the estimated effective penetration depth $\Lambda_{ab} = \lambda_{ab}^2/t \sim 1.15$ μm (we used the bulk YBCO $\lambda_{ab} = 150$ and $t = 20$ nm is the film thickness), from which one could actually expect $U_m$ to be sensibly smaller than $\sim \phi_0 \langle |H_S| \rangle$.

In conclusion, we have shown that the stray magnetic fields from a ferromagnet can strongly modify the mixed-state magneto-transport of a High-$T_C$ superconductor. We found a magnetic-history-controlled negative/positive resistance switching, reminiscent of the standard/inverse "spin switch effects" [10-20], which is produced here by vortex dynamics and pinning effects. Our results shows that, in order to optimize magnetic pinning in high-$T_C$'s, the F domain structure sizes must be chosen in order to limit the drawback of the dissipation caused by the wandering of the vortices/antivortices induced by the stray field.

This work was supported by the French RTRA "Supraspin" and ANR "SUPERHYBRIDS-II" grants.

[30] The multilayer was treated as a single magnetic layer with a reduced effective $M_S = t_{Co}/t_{ml} M_{Co}$, with $M_{Co} = 1.446$ emu cm-2 the Co saturation magnetization and $t_{Co}/t_{ml} = 0.375$ the ratio between the total thickness of the Co layers and the full multilayer. MFM images were converted to binary and each pixel treated as a uniformly magnetized region (perpendicular for states II and III, ~20º off the c-axis for state I, as implied by AHE, Fig. 2 (d)). The field due to the resultant surface magnetic charges on each pixel were summed together to yield $H_S(x)$ (calculated at the center of each pixel and in the center of the YBCO layer).

**FIGURE CAPTIONS**

**Figure 1:** Resistance *vs.* applied field $H$ as swept from positive to negative (black) and *vice versa* (red dashed), for different angles between $H$ and the YBCO c-axis (see sketch and legends). Open blue circles are for the virgin curves, measured after demagnetization. (a) for the S/F sample $T=0.85T_C$ and (b) for a single 20 nm thick YBCO film at $T=0.72T_C$, in both cases with J=37.5 kA cm$^{-2}$.

**Figure 2:** (a)-(d), each figure for a different angle $\theta$ between the applied field and the YBCO c-axis (see legends). Top panels: absolute resistance switching (left axis) and percent resistance switching with respect to the background magneto-resistance (right axis) as a function of the applied field H. For $T=0.89T_C$ and J= J=37.5 kA cm$^{-2}$. Bottom panels: Net perpendicular component of the multilayer's magnetization as a function of the applied field at 100 K. The sketches are schematics of the magnetic states at the cycle points indicated by the arrows.

**Figure 3:** (a) $E(J)$ for different temperatures (see labels) in a field $H=1$ kOe ($\theta=80°$) after application of 3.5 kOe (black solid) and –3.5 kOe (red dashed). The inset shows a zoom of the curve for 60 K. (b) percent $\Delta R(J)$ obtained from the curves in (a). Inset: enhancement of the critical current induced by the domain structure as a function of $H$ ($\theta=80°$), for different temperatures temperatures 62 K, 60 K and 57 K. (c) $E(J)$ for different temperatures (see labels) in an applied field ($\theta=90°$) $H=-4$ kOe (black solid) and $H=0.5$ kOe (red dashed). (d) percent $\Delta R(J)$ obtained from the curves in (c).

**Figure 4:** (a)-(c) room-temperature MFM of the remanent domain structure after I) application and withdrawal of a saturating in-plane field and II) in-plane and III) out-of-plane demagnetization. (d)-(f) profile of the perpendicular component of the stray magnetic field $H_S$ from the domain structures I, II, and III . (g) Zero-field (remanent) resistance for different magnetic states (uniform out-of-plane

magnetization, I, II, and III) plotted as function of the magnetic flux quanta $\phi/\phi_0$ generated by the average magnetic domain, for different temperatures.

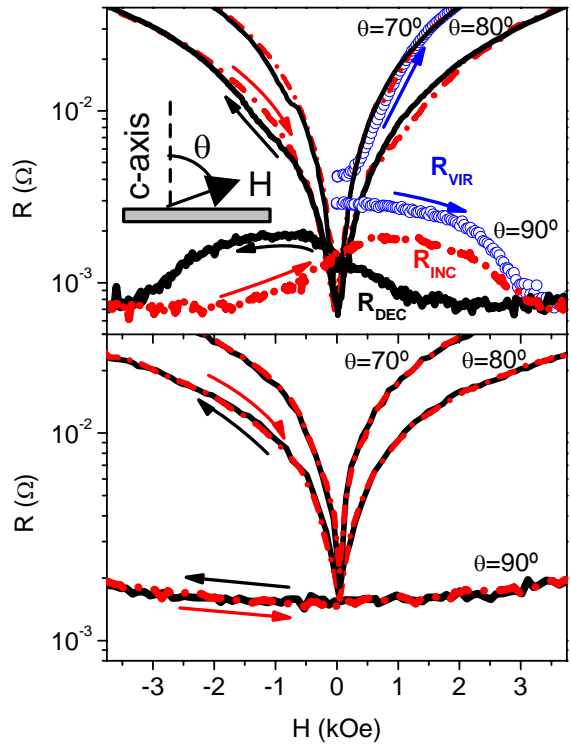

Fig. 1

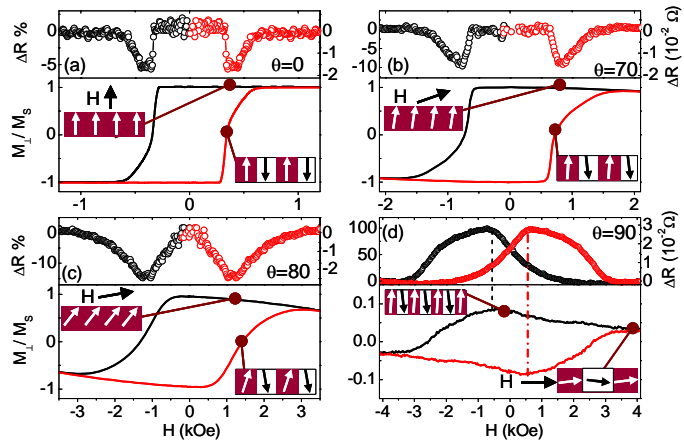

Fig. 2

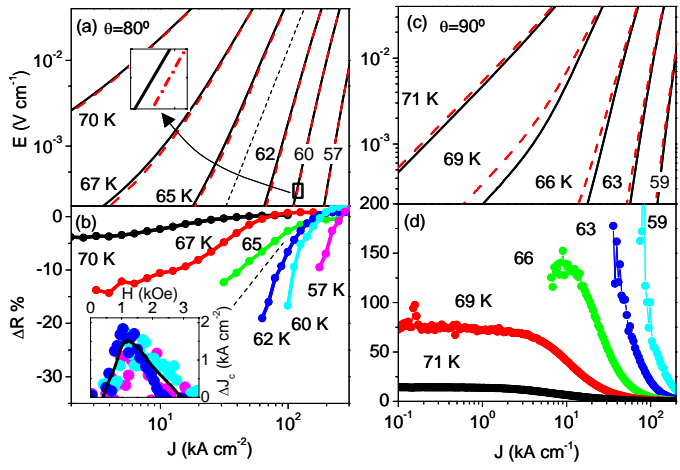

Fig. 3

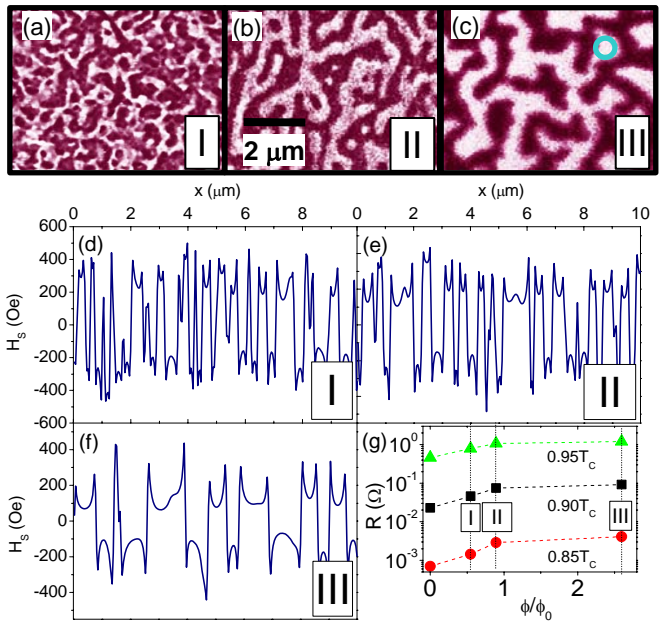

Fig. 4.